\documentclass[letterpaper]{article} 
\usepackage{aaai2026}  
\usepackage{times}  
\usepackage{helvet}  
\usepackage{courier}  
\usepackage[hyphens]{url}  
\usepackage{graphicx} 
\usepackage{natbib}  
\usepackage{caption} 
\usepackage{algorithm}
\usepackage{algorithmic}

\usepackage{amssymb}
\usepackage{amsmath}
\usepackage{subfigure}
\usepackage{booktabs}
\usepackage{multirow}
\usepackage{array}
\usepackage{amsthm}

\usepackage{newfloat}
\usepackage{listings}
\DeclareCaptionStyle{ruled}{labelfont=normalfont,labelsep=colon,strut=off} 
\floatstyle{ruled}
\newfloat{listing}{tb}{lst}{}
\floatname{listing}{Listing}
\title{CellStream: Dynamical Optimal Transport Informed Embeddings for Reconstructing Cellular Trajectories from Snapshots Data}
\author{
    Yue Ling\textsuperscript{\rm 1}\equalcontrib, 
    Peiqi Zhang\textsuperscript{\rm {1,5}}\equalcontrib, 
    Zhenyi Zhang\textsuperscript{\rm 1}\thanks{Corresponding authors.}, 
    Peijie Zhou\textsuperscript{\rm {2,3,4,5}\textdagger}
}
\affiliations{
    \textsuperscript{\rm 1}School of Mathematical Sciences, Peking University, China\\

    \textsuperscript{\rm 2} Center for Machine Learning Research, Peking University, China
\\ \textsuperscript{\rm 3} Center for Quantitative Biology, Peking University, China \\
    \textsuperscript{\rm 4} National Engineering Laboratory for Big Data Analysis and Applications, China \\
    \textsuperscript{\rm 5} AI for Science Institute, Beijing, China
    \\
    zhenyizhang@stu.pku.edu.cn; pjzhou@pku.edu.cn
}

\begin{document}

\maketitle

\begin{abstract}
Single-cell RNA sequencing (scRNA-seq), especially temporally resolved datasets, enables genome-wide profiling of gene expression dynamics at single-cell resolution across discrete time points. However, current technologies provide only sparse, static snapshots of cell states and are inherently influenced by technical noise, complicating the inference and representation of continuous transcriptional dynamics. Although embedding methods can reduce dimensionality and mitigate technical noise, the majority of existing approaches typically treat trajectory inference separately from embedding construction, often neglecting temporal structure.  To address this challenge, here we introduce CellStream, a novel deep learning framework that jointly learns embedding and cellular dynamics from single-cell snapshots data by integrating an autoencoder with unbalanced dynamical optimal transport. Compared to existing methods, CellStream generates dynamics-informed embeddings that robustly capture temporal developmental processes while maintaining high consistency with the underlying data manifold. We demonstrate CellStream’s effectiveness on both simulated datasets and real scRNA-seq data, including spatial transcriptomics. Our experiments indicate significant quantitative improvements over state-of-the-art methods in representing cellular trajectories with enhanced temporal coherence and reduced noise sensitivity. Overall, CellStream provides a new tool for learning and representing continuous streams from the noisy, static snapshots of single-cell gene expression.
\end{abstract} 


\begin{links}
     \link{Code}{https://github.com/PQ-Zhang/CellStream}
\end{links}
\section{Introduction}
Time-resolved single-cell RNA sequencing (scRNA-seq) captures gene expression at the genome scale in individual cells across discrete time points, enabling analysis of dynamic biological processes. However, because sequencing destroys cells, it yields sparse, noisy snapshots that obscure continuous dynamical trajectories~\cite{waddingot,moscot,review_OT}. Dimensionality reduction through low-dimensional embeddings can partially alleviate sparsity and noise issues, and provide insights of the temporal inference through pseudotime analysis~\cite{review_OT}. However, existing embedding methods predominantly focus on preserving topological or geometric relationships~\cite{lowD1,lowD2,lowD3}. They often neglect the temporal structure embedded in time-series snapshots data, potentially resulting in misinterpretation of cellular trajectories~\cite{zheng2023pumping,chari2023specious}. Thus, developing embeddings that explicitly incorporate temporal structure to faithfully reconstruct continuous cellular dynamics remains an open and critical challenge in the field.

For a single snapshot scRNA-seq data, many embedding methods are proposed to model and visualize the biological processes on a low-dimensional manifold. For example, embeddings derived from principal component analysis (PCA), t-distributed Stochastic Neighbor Embeddings (t-SNE), Uniform Manifold Approximation and Projection (UMAP) and diffusion maps are typical methods to decipher the topological structure of gene expression~\cite{tSNE,UMAP,Diffusionmaps}. Canonical correlation analysis (CCA), Harmony and Partition-based graph abstraction (PAGA) shed light on preserving the statistical, geometric or topological structure~\cite{CCA,Harmony,PAGA}. However, these embedding methods usually focus on retaining the geometric structure but are not designed to capture the cell-state transition dynamics and temporal relationships between snapshots. To address the problem, CellPath, Ocelli and VeloViz propose using RNA velocity results \cite{RNAvelocity} to conduct velocity-informed visualization and pseudo-time trajectory inference~\cite{CellPath,Ocelli,VeloViz}. However, the RNA velocity inference could encounter computational issues \cite{RNAvelocitydrawback,zheng2023pumping}, especially in datasets with low unspliced counts, and does not consider cross-time point analysis explicitly. While deep generative model embeddings, like scVI, Geneformer and TarDis \cite{scVI,Geneformer,TarDis}, attempt to address batch effects and library size effects, they suffer from limited interpretability for the dynamical modeling.

To deal with the time-series scRNA-seq data, Optimal transport (OT) has emerged as a promising tool to model the cellular dynamics ~\cite{bunne2024optimal,review_OT}. Its core idea relies on seeking the optimal transport plan between the population of cells among consecutive time points with minimal transportation cost~\cite{waddingot,moscot}. To deal with cell proliferation effects, TIGON proposes the application of unbalanced dynamical OT in trajectory inference~\cite{TIGON}. Subsequent works further relate the regularized unbalanced optimal transport (RUOT) and mean-field terms to model stochastic effects and cell-cell communications, proposing a new framework with higher performance and efficiency~\cite{DeepRUOT,DeepRUOTv2,VarRUOT}. Nevertheless, since direct inference of cellular dynamics in a high-dimensional gene expression space is usually unstable and demands high computational resources~\cite{OTinhighD}, these methods usually rely on a pre-defined embedding as the input such as PCA or UMAP space, which is decoupled from the dynamical learning.

To address these challenges, we present CellStream, a deep learning framework that generates dynamics-informed embeddings based on unbalanced dynamical OT for time-series single-cell snapshots data. We evaluate the effectiveness of CellStream in modeling cellular dynamics using both simulated data and real data. Experiments demonstrate that CellStream embeddings can precisely reveal developmental trajectories from the noisy, static snapshots of single-cell gene expression. Overall, our method features the following contributions: 
\begin{itemize}
    \item We introduce CellStream that \textbf{jointly learns embedding and latent cellular dynamics} from cross-sectional snapshots of single-cell data by integrating an autoencoder with unbalanced dynamical optimal transport;
    \item We demonstrate that CellStream embeddings are capable of reconstructing the cellular dynamics \textbf{robustly against noise} through simulation benchmarking.
    \item We validate the effectiveness of CellStream on both simulated data and real snapshots data, showing its significant quantitative improvements
over state-of-the-art methods in \textbf{representing cellular trajectories with enhanced temporal coherence}.
\end{itemize}

\begin{figure}[!htb]
\centering
\includegraphics[width=1.0\linewidth]{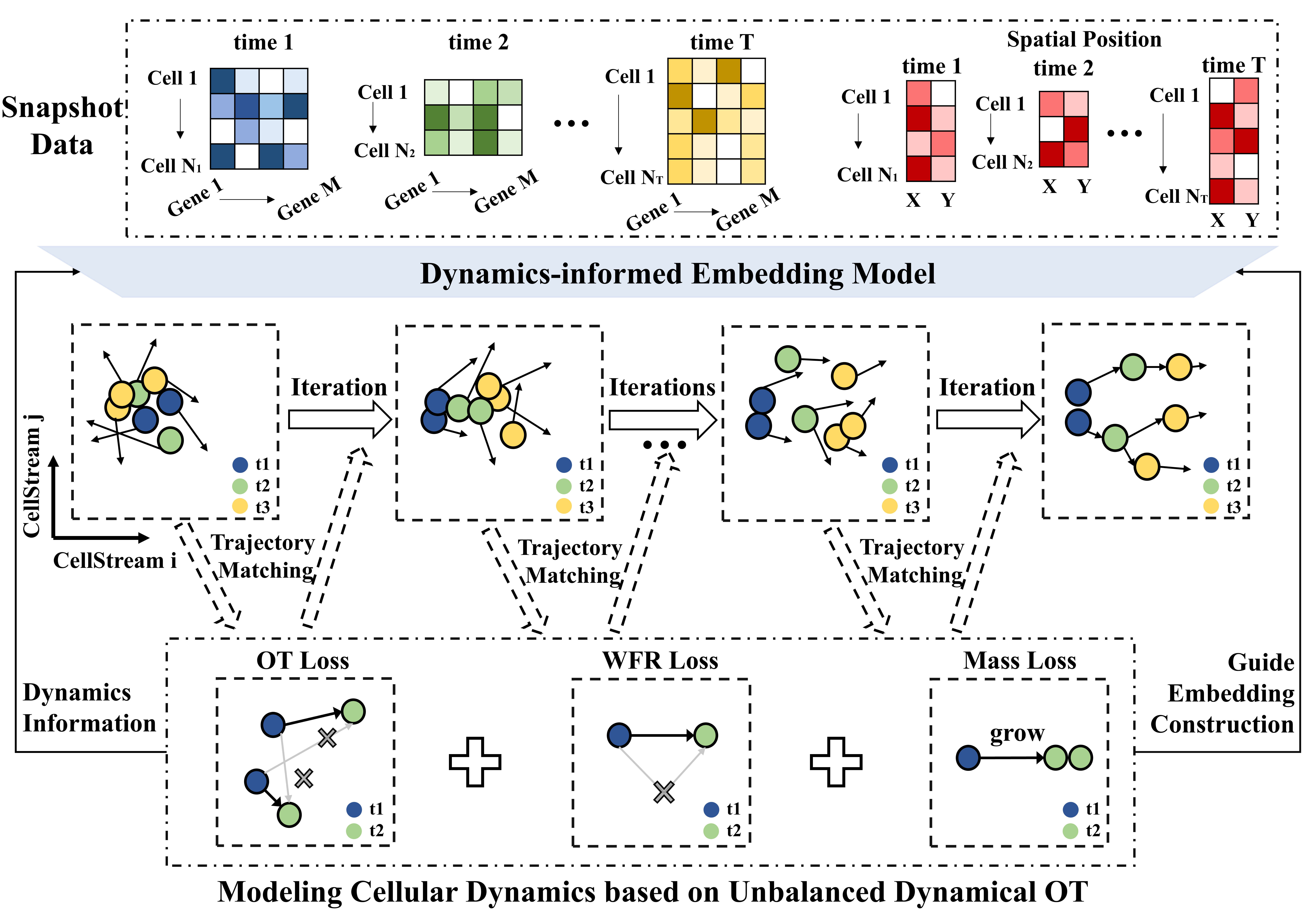}
\caption{ Overview of CellStream. CellStream is a dynamics-informed embedding model that jointly learns embeddings and continuous cellular streams from static, sparse snapshots data. CellStream uses an autoencoder and neural network to learn embeddings and continuous cellular streams from sparse snapshots data, guided by real-time trajectory feedback composed of $\mathcal{L}_{\text{OT}}$, $\mathcal{L}_{\text{WFR}}$ and $\mathcal{L}_{\text{Mass}}$.}
\label{fig:framework}
\vspace{-0.1cm}
\end{figure}

\section{Related Work}
\subsection{Manifold-learning Embedding} 
Numerous methods have been developed to project the complex, high-dimensional data onto low-dimensional manifolds. PCA, t-SNE and UMAP are widely used to reveal the spatial distribution of the given cellular trajectories~\cite{TIVelo,UniTVelo,DeepRUOT}. Diffusion map models the geometry of the dataset as a diffusion process on point clouds, capturing local similarity and global structure via the eigen-space of Markov chains~\cite{Diffusionmaps}. Canonical correlation analysis (CCA) constructs the embedding space by identifying linear transformations that maximize the correlations between multiple datasets to mitigate technical  noise and preserve biological signals~\cite{CCA,CCAapplication}. Similarly,  Harmony performs iterative soft clustering and linear corrections in the latent space to remove batch effects~\cite{Harmony,Harmonyapplication}. Partition-based graph abstraction (PAGA) builds a topology-preserving embedding space through graph abstraction, coarsening single-cell neighborhood graphs to retain geometric structure~\cite{PAGA,PAGAapplication}. Although these approaches specialize in extracting topological traits from individual snapshots, they tend to overlook dynamic information and not fully consider temporal structure across different time points.

\subsection{Dynamical Embedding}
In recent years, several studies have explored how to manifest cellular dynamics in the embedding space. By integrating multiscale stochastic dynamics, Multiscale Transition Analysis (MuTrans) generates a dynamical manifold embedding that maps stable cell states to attractor basins and transient cells to saddle points, revealing continuous cell fate trajectories~\cite{Mutrans}. CellPath constructs meta-cell level embeddings based on gene expression similarity and RNA velocity coherence, employing a directed KNN graph for pseudo-time trajectory inference through path finding algorithms~\cite{CellPath}. Integrating RNA velocity and multi-modal diffusion maps, Ocelli represents the cellular differentiation direction on a low-dimensional embedding from multimodal single-cell data~\cite{Ocelli}. VeloViz constructs RNA velocity-informed embeddings based on a weighted k-nearest neighbor graph which integrates RNA velocity with gene expression~\cite{VeloViz}. However, the steady-state assumption of MuTrans may limit its application to non-stationary systems with rapid proliferation, and RNA velocity used in CellPath, Ocelli and VeloViz faces challenges to model complex temporal transitions across time points which have longer time scales than splicing dynamics~\cite{zhou2024spatial}. 

\subsection{Deep Generative Model Embedding}
Deep generative model embeddings leverage deep learning to generate robust, low-dimensional representations, enabling batch correction and regulatory inference. By integrating variational inference and conditional generative, scVI provides low-dimensional embeddings free from batch effects and library size effects~\cite{scVI}. GeneFormer employs a transformer architecture to tokenize rank-ordered gene expressions into sequences for self-supervised masked prediction, generating contextual cell embeddings that encode regulatory network dynamics~\cite{Geneformer}. TarDis constructs embeddings in a variational autoencoder framework and leverages self-supervised disentanglement losses to distinguish covariate-specific variations from underlying biological processes~\cite{TarDis}. However, the majority of existing deep generative model embeddings are usually established on VAE or Transformer architectures,  encountering challenges in the interpretability issue of dynamical modeling.

\subsection{Dynamical OT based Trajectory Inference}
Dynamical OT proves to be an effective model for deciphering cellular dynamics~\cite{review_OT,gu2025partially}. TrajectoryNet pioneers the application of dynamical OT in trajectory inference combined with continuous normalizing flows~\cite{TrajectoryNet}. MIOFlow introduces a Geodesic Autoencoder (GAE) to learn low-dimensional manifolds which enable trajectory inference from time-series snapshots based on dynamical optimal transport theory~\cite{MIOFlow}. Leveraging unbalanced dynamical OT, TIGON infers differentiation trajectories and population growth from scRNA-seq data simultaneously~\cite{TIGON}. stVCR further reconstructs the interaction between gene expression and spatial migration from spatial transcriptome (ST) data~\cite{stvcr,zhang2025deciphering}. Focused on intercellular communications, CytoBridge relates the regularized unbalanced optimal transport (RUOT) to Mean-Field Schrödinger Bridge (MFSB), recovering unbalanced stochastic interaction dynamics~\cite{DeepRUOT,DeepRUOTv2,VarRUOT}. There are also many flow matching-based methods to tackle this problem \citep{flowmatching3,flowmatching4,flowmatching6,flowmatching7,wang2025joint} and also some methods are based on the Schrödinger Bridge (SB) problem \citep{gwot,bunne_unsb}. However, to tackle sparsity and noise problems in original gene expression space, existing works often rely on a reduced dimension input as pre-computed embeddings.  The separation of trajectory inference from embedding construction is limited when considering the temporal structure of datasets, impeding the reconstruction of spatiotemporal dynamics.

\section{Methodology}

\subsection{Optimal Transport}
\noindent\textbf{Static Optimal Transport.}
Optimal transport(OT) seeks to minimize the transport cost between two populations, namely the Kantorovich problem~\cite{Kantorovich}. 
For $p \in [1, +\infty]$, the Wasserstein $p$-distance is defined as
\begin{equation*}
    \mathcal{W}_{p}(\mu, \nu) = \inf_{\gamma \in \Gamma (\mu , \nu)} \left(\mathbb{E}_{(\mathbf{x},\mathbf{y})\sim \gamma} \left[ d(\mathbf{x},\mathbf{y})^p \right] \right)^{1/p},
\end{equation*}
where $\mu \in \mathcal{P}(X)$ and $\nu \in \mathcal{P}(Y)$, and a coupling $\gamma$ is a joint probability measure whose marginals are $\mu$ and $\nu$. $\Gamma(\mu, \nu)$ denotes the set of all couplings and $d : X \times Y \rightarrow \mathbb{R}^+$ is a given distance function. $\mathcal{W}_{p}(\mu, \nu)$ represents the optimal transport cost between two probability measures.  Unless explicitly stated, we choose $p=2$  by default in the calculation.

\noindent\textbf{Dynamical Optimal Transport.}
Equivalent to static OT, the framework of dynamical OT was formulated in the form of fluid dynamics equations, which modeled the transport as a continuous flow and optimized over smooth and time-dependent density $\rho(t, \mathbf{x})$ and velocity fields $\mathbf{v}(t, \mathbf{x}) $~\cite{dynamicalOT}. The spatiotemporal dynamics of the mass distribution was subject to the following continuity equation with the initial and final constraints:
\begin{equation*}
    \label{eq: dOT}
    \partial_t\rho + \nabla_{\mathbf{x}}\cdot(\mathbf{v}\rho) = 0, \quad 
    \rho(0, \cdot) = \rho_0,\quad \rho(T, \cdot) = \rho_T
\end{equation*}
for any $t \in [0,T]$ and $\mathbf{x}\in\mathbb{R}^d$, 
where 
$\mathbf{v}(t, \mathbf{x})\in\mathbb{R}^d$
determined the velocity field that governed the dynamics of the density. When defining the transport distance function as the  Euclidean distance, namely $d(\mathbf{x},\mathbf{y}) = {\|\mathbf{x}-\mathbf{y}\|}_{2}$,  the cost function of dynamical OT can be expressed as
\begin{equation*}
        \mathcal{L}_{\text{dynamical OT}} = \int_{0}^{T}\int_{\mathbb{R}^d} {{\|\mathbf{v}(t, \mathbf{x})\|}^2} \rho(t,\mathbf{x}) \, \mathrm{d}\mathbf{x} \mathrm{d}t.
\end{equation*}

\noindent\textbf{Unbalanced Dynamical Optimal Transport.}
The mass conservation assumption of dynamical OT does not account for the total mass changing, which is an important phenomenon in biological processes due to cell apoptosis or divide. To model such dynamics, unbalanced dynamical OT introduces a growth term 
$g(t, \mathbf{x}):{ \left[ 0,T \right] \times \mathbb{R}^d}\rightarrow\mathbb{R}$
to rewrite the continuity equation as
\begin{equation*}
   \partial_t \rho + \nabla_{\mathbf{x}} \cdot (\mathbf{v} \rho) = g \rho, \quad
    \rho(0, \cdot) = \rho_0, \quad 
\rho(T, \cdot) = \rho_T.
\end{equation*}
A transport cost function that considers both kinetic and growth energy simultaneously is imperative to constrain the transport dynamics. Wasserstein–Fisher–Rao (WFR) distance emerges as a desirable choice to describe the overall shipping cost~\cite{WFR1, WFR2}, and the corresponding cost function of unbalanced dynamical OT can be defined as:
\begin{equation}
    \label{eq: WFR}
    \mathcal{L}_{\text{WFR}} =  \int_0^T \int_{\mathbb{R}^d} \left.(\|\mathbf{v}(t, \mathbf{x})\|^2 +\\
    \alpha g^2(t, \mathbf{x}) \right.) 
    \rho(t,\mathbf{x})\, \mathrm{d}\mathbf{x}\, \mathrm{d}t,
\end{equation}
where integral terms involving $\|\mathbf{v}(t, \mathbf{x})\|^2$ and $g^2(t, \mathbf{x})$ are used to quantify the cost of transport and growth respectively. The hyperparameter $\alpha$ is introduced to explicitly control the relative weight of the above two terms when minimizing the cost function, and we set $\alpha=1$ in our tasks. The sensitivity of our algorithm to $\alpha$ is presented in Appendix B.5.

\subsection{Modeling the Single Cell Transcriptional Dynamics in the Low-dimensional Embedding Space}

Despite the high-dimensionality of gene expression, it is commonly assumed that the dynamical biological processes underlying scRNA-seq data are indeed embedded on a low-dimensional manifold ~\cite{lowD1, lowD2, lowD3}.  CellStream aims to model the dynamics of cellular trajectories and growth in an iteratively refined low-dimensional embedding space based on unbalanced dynamical OT. 
Formally, our goal is to jointly learn a suitable time-dependent low-dimensional embedding $f^{\text{enc}}(t, \mathbf{x})$, which maps the original $D$-dimensional data with time $t$ to a $d$-dimensional space($d < D$), together with the dynamical functions $\mathbf{v}$ and $g$ defined in this embedding space, such that the WFR distance $\mathcal{L}_{\text{WFR}}^{\text{emb}}$ in the dynamical optimal transport problem is minimized.

The constraint of $f^{\text{enc}}(t, \mathbf{x})$ is enforced by introducing an embedding loss $\mathcal{L}_{\text{embedding}}$, which promotes faithful reconstruction of the input in the latent space.
For example, if $f^{\text{enc}}$ is implemented as the encoder module of an autoencoder, a decoder 
$f^{\text{dec}}: \mathbb{R}^d \rightarrow \mathbb{R} \times \mathbb{R}^D$
can be jointly trained, and the embedding loss is defined as
$\mathcal{L}_{\text{embedding}} = \text{MSE}\left( f^{\text{dec}} \circ f^{\text{enc}}([t, \mathbf{x}]),\, [t, \mathbf{x}] \right)$.
The specific form of the problem is as follows:
\begin{equation*}
    \operatorname*{min}_{f^{\text{enc}}, f^{\text{dec}}, \mathbf{v}, g} \, \mathcal{L}_{\text{embedding}}(f^{\text{enc}}, f^{\text{dec}}) + \lambda \, \mathcal{L}_{\text{WFR}}^{\text{emb}}(f^{\text{enc}}, \mathbf{v}, g)
\end{equation*}
where
$f^\text{enc}(t, \mathbf{x}): \mathbb{R} \times \mathbb{R}^D \rightarrow \mathbb{R}^d$, 
$\mathbf{v}(t, \mathbf{z}): \mathbb{R} \times \mathbb{R}^d \rightarrow \mathbb{R}^d$, 
$g(t, \mathbf{z}): \mathbb{R} \times \mathbb{R}^d \rightarrow \mathbb{R}$, 
with $\mathbf{z} = f^{\text{enc}}(t, \mathbf{x})$.

The system is constrained by
$\rho(0, \cdot) = \rho_0$, $\rho(T, \cdot) = \rho_T$, 
and satisfies the continuity equation
\begin{equation}
\label{eq: continuity equation}
    \partial_t q(t, \mathbf{z}) + \nabla_{\mathbf{z}} \cdot (\mathbf{v}(t, \mathbf{z}) q(t, \mathbf{z})) = g(t, \mathbf{z}) q(t, \mathbf{z}),
\end{equation}
where $q = {f_{\#}^{\text{enc}}} \rho$ , i.e. $q(t, \mathbf{z}) = \int \rho(t, \mathbf{x}) \delta(\mathbf{z} - f^{\text{enc}}(\mathbf{x})) \mathrm{d} \mathbf{x}$.

\subsection{Deep Learning-based Solver in CellStream}

We suppose that time series RNA-seq data are given by
\begin{equation*}
\begin{split}
(t_i, \mathbf{X}_i)_{i=0}^{T-1},\quad
\mathbf{X}_i = \{ \mathbf{x}_{i}^j \in \mathbb{R}^D\}_{j=1}^{N_i}
\end{split}
\end{equation*}
where $D$ indicates the dimension of gene expression, $N_i$ represents the cell population at time $t_i$ and we use index $j$ for the single cells. We assume that the number of cells involved at time $t_i$ is proportional to $N_i$ if there is no prior information.
In the embedding space, the input data is reduced to 
\begin{equation*}
(t_i, \mathbf{Z}_i, \mathbf{w}_i)_{i=0}^{T-1},\quad
\mathbf{Z}_i = \{ \mathbf{z}_{i}^j = f^\text{enc}(\mathbf{x}_{i}^j) \in \mathbb{R}^d\}_{j=1}^{N_i}
\end{equation*}
where $\mathbf{w}_i = \frac{1}{N_i}\mathbf{1}_{N_i} \in\mathbb{R}^{N_i}$ implies the weight of each point. Driven by $\mathbf{v}$ and $g$, we can also obtain predicted states:
\begin{equation*}
    \left(t_i, \hat{\mathbf{Z}}_i,\, \hat{\mathbf{w}}_i = \{\hat{w}^j(t_i)\}_{j=1}^{N_0} \right)_{i=0}^{T-1},\,
    \hat{\mathbf{Z}}_i = \{ \hat{\mathbf{z}}^j(t_i) \in \mathbb{R}^d \}_{j=1}^{N_0}
\end{equation*}
initialized from
$\hat{\mathbf{Z}}_0 = \mathbf{Z}_0$, 
$\hat{\mathbf{w}}_0 = \mathbf{w}_0$
by ODE such that 
\begin{equation}
    \label{eq: evolution ODE}
    \mathrm{d}\hat{\mathbf{z}}^j(t) / \mathrm{d}t = 
    \mathbf{v} (t, \hat{\mathbf{z}}^j), \quad
    \mathrm{d}\text{log}\hat{w}^j(t) / \mathrm{d}t = 
    g(t, \hat{\mathbf{z}}^j).
\end{equation}
In order to construct a feasible and efficient solver for CellStream, we implement $f^{\text{enc}}$ as an autoencoder and parameterize these functions to be solved (including autoencoder $f^{\text{enc}}$ and $f^{\text{dec}}$, velocity field $\mathbf{v}$, and growth term $g$) based on neural networks as $f_\theta, \mathbf{v}_\phi, g_\psi$. Here $\theta, \phi, \psi$ are parameters to be optimized in neural networks. The total loss function of neural networks ($\mathcal{L}$) can be divided into autoencoder loss ($\mathcal{L}_{\text{AE}}$), WFR distance loss ($\mathcal{L}_{\text{WFR}}$), and data matching loss ($\mathcal{L}_{\text{Match}}$):
\begin{equation}
    \label{eq: loss function}
    \mathcal{L} = \lambda_{\text{AE}} \mathcal{L}_{\text{AE}} + \lambda_{\text{WFR}} \mathcal{L}_{\text{WFR}}^{\text{emb}} + \lambda_{\text{Match}} \mathcal{L}_{\text{Match}}.
\end{equation}
The three loss terms refer to autoencoder reconstruction error, optimal transport distance, and matching error. $\lambda_{\text{AE}}, \lambda_{\text{WFR}}$ and $\lambda_{\text{Match}}$ are hyperparameters controlling the relative weight between loss terms. In the following experiments, we set $\lambda_{\text{AE}}=10,\, \lambda_{\text{WFR}}=1,\, \lambda_{\text{Match}}=5$. We will verify the robustness of the algorithm to those parameters in Appendix B.5.

\noindent\textbf{Autoencoder Loss.} To minimize the reconstruction error between the estimated single-cell data and the observed measurements, we introduce autoencoder loss $\mathcal{L}_{\text{AE}}$ based on Mean Squared Error (MSE) :
\begin{equation*}
    \mathcal{L}_{\text{AE}}(\theta) = \text{MSE}( f_\theta^{\text{dec}} \circ f_\theta^{\text{enc}}(t, \mathbf{x}),\, [t, \mathbf{x}] )
\end{equation*}

\noindent\textbf{Wasserstein-Fisher-Rao  Loss in Embedding Space.} We can rewrite \eqref{eq: WFR} under the circumstances of neural networks as:
\begin{equation}
\begin{split}
    \label{eq: WFR_nn}
    \mathcal{L}_{\text{WFR}}^{\text{emb}}(\theta, \phi, \psi) &= \int_0^T \int_{\mathbb{R}^d} \left( \|\mathbf{v}_\phi(t, \mathbf{z_\theta})\|^2 \right. \\
    &\qquad + \left. \alpha g_\psi^2(t, \mathbf{z_\theta}) \right) q(t, \mathbf{z_\theta}) \, \mathrm{d}\mathbf{z_\theta}\, \mathrm{d}t.
\end{split}
\end{equation}
The integral form in \eqref{eq: WFR_nn} involves computing the density term $q(t,\mathbf{z})$ that satisfies \eqref{eq: continuity equation}, which is computationally prohibitive in practice since it requires solving complex partial differential equations (PDE) in high-dimensional space. To address this computational bottleneck, we leverage the fundamental equivalence between Eulerian and Lagrangian descriptions of fluid dynamics and  further convert \eqref{eq: WFR_nn} to an equivalent dimensionless form~\cite{TIGON, DeepRUOT}:
\begin{equation*}
\begin{split}
    \label{eq: WFR_E}  
    \mathcal{L}_{\text{WFR}}^{\text{emb, par}}(\theta, \phi, \psi) &= \sum_{j=1}^{N_0}
    \int_0^T \left( \|\mathbf{v}_\phi(t, \hat{\mathbf{z}}^j(t))\|^2 \right. \\
    &\qquad + \left. \alpha g_\psi^2(t, \hat{\mathbf{z}}^j(t)) \right) 
    \hat{w}^j(t)\, \mathrm{d}t,
\end{split}
\end{equation*}
where $\hat{\mathbf{z}}^j(t)$ can be calculated efficiently using Neural ODE solver \cite{chen2018neural} for \eqref{eq: evolution ODE}.

\noindent\textbf{Data Matching Loss.} Data matching loss aims to match the distribution of the estimated data at each time point with its corresponding ground truth. The Wasserstein distance is widely used to quantify such distribution difference. However, since we relax the mass conservation assumption to the unbalanced event \eqref{eq: continuity equation}, the combination of Mass Loss $\mathcal{L}_{\text{Mass}}$ and OT Loss $\mathcal{L}_{\text{OT}}$ is calculated as
$ \mathcal{L}_{\text{Match}}(\theta, \phi, \psi) = \lambda_{\text{Mass}}\mathcal{L}_{\text{Mass}} + \lambda_{\text{OT}}\mathcal{L}_{\text{OT}},$
where $\mathcal{L}_{\text{Mass}}$ aims to ensure the consistency of cell population and $\mathcal{L}_{\text{OT}}$ is used to evaluate the gap between different distributions. $\lambda_{\text{Mass}}$ and $\lambda_{\text{OT}}$ are hyperparameters to balance the relative weight, and we set $\lambda_{\text{Mass}}=1, \,\lambda_{\text{OT}}=1$ in the following experiments.

Data matching loss $\mathcal{L}_{{\text{Match}}}$ is intended to quantify the differences between the estimated distribution $(t_i, \hat{\mathbf{Z}}_i, \hat{\mathbf{w}}_i)_{i=0}^{T-1}$ and the ground truth $(t_i, \mathbf{Z}_i, \mathbf{w}_i)_{i=0}^{T-1}$. 
For Mass Loss $\mathcal{L}_{\text{Mass}}$, we use the global mass difference as the measurement:
\begin{equation*}
    \mathcal{L}_{\text{Mass}}(\theta, \phi, \psi) = \sum_{i=0}^{T-1}
    \left| \sum_{j=0}^{N_0-1} \hat{w}_i^j - \frac{N_i}{N_0} \right|.
\end{equation*}
For OT Loss $\mathcal{L}_{\text{OT}}$, we approximate it by using the Wasserstein 2-distance between two normalized distributions:
\begin{equation*}
    \mathcal{L}_{\text{OT}}(\theta, \phi, \psi) = \sum_{i=0}^{T-1}
    \mathcal{W}_2\left(
    \cfrac{\hat{\mathbf{w}_i}}{\|\hat{\mathbf{w}}_i\|_1}, 
    \cfrac{\mathbf{w}_i}{\|\mathbf{w}_i\|_1}
    \right).
\end{equation*}

\subsection{Iterative Training Strategies}

To improve the robustness and efficiency of convergence, we initialize the autoencoder using PCA. In practice, the encoder $f_\theta^{\text{enc}}$ and decoder $f_\theta^{\text{dec}}$ are optimized separately to minimize the loss function of the autoencoder $\mathcal{L}_{\text{AE}}$. 

Due to the large number of parameters and the intricate calculation of loss functions, we employ Block Coordinate Descent (BCD), which involves separating the autoencoder from the dynamic components and alternately optimizing the parameters of both~\cite{BCD}. Instead of training them independently, when training the autoencoder to obtain cell embeddings, we not only take into account $\mathcal{L}_{\text{AE}}$ but also incorporate the dynamics loss function (including $\mathcal{L}_{\text{WFR}}, 
\mathcal{L}_{\text{OT}}$ and $\mathcal{L}_{\text{Mass}}$ ) proportionally. These training strategies facilitate the development of an interpretable embedding space. The specific training strategies are as follows:

\begin{algorithm}
\caption{Training CellStream}
\begin{algorithmic}
\STATE \textbf{Input}: Time Series scRNA data $(t_i, \mathbf{X}_i)_{i=0}^{T-1}$, the dimension of embedding space $d$, maximum iterations $n_\text{max}$.
\STATE \textbf{Initialization}:$\theta$, $\phi$, $\psi$
\WHILE{the loss function remains decreasing}
    \STATE \textbf{Optimize} $\theta$ \textbf{when fixing} $\phi$, $\psi$ \textbf{:}
    \STATE Calculate $(t_i, \mathbf{Z}_i, \mathbf{w}_i)_{i=0}^{T-1}$ based on the input and $f_\theta^{\text{enc}}$
    \FOR{$\text{iter} = 1$ to $n_\text{max}$}
        \STATE Estimate $(t_i, \hat{\mathbf{Z}}_i, \hat{\mathbf{w}}_i)_{i=0}^{T-1}$ via NeuralODE using $(t_0, \mathbf{Z}_0)$ by \eqref{eq: evolution ODE}
        \STATE Calculate the loss function: $\mathcal{L} \gets \lambda_{\text{AE}} \mathcal{L}_{\text{AE}} + \lambda_{\text{WFR}} \mathcal{L}_{\text{WFR}}^{\text{emb, par}} + \lambda_{\text{Match}} \mathcal{L}_{\text{Match}}$
        \STATE Update $\theta$
    \ENDFOR
    \STATE \textbf{Optimize} $\phi$, $\psi$ \textbf{when fixing} $\theta$ \textbf{:}
    \STATE Calculate$(t_i, \mathbf{Z}_i, \mathbf{w}_i)_{i=0}^{T-1}$ based on the input and $f_\theta^{\text{enc}}$
    \FOR{$\text{iter} = 1$ to $n_\text{max}$}
        \STATE Estimate $(t_i, \hat{\mathbf{Z}}_i, \hat{\mathbf{w}}_i)_{i=0}^{T-1}$ via Neural ODE using $(t_0, \mathbf{Z}_0)$ by \eqref{eq: evolution ODE}
        \STATE Calculate the loss function: $\mathcal{L} \gets \lambda_{\text{WFR}} \mathcal{L}_{\text{WFR}}^{\text{emb, par}} + \lambda_{\text{Match}} \mathcal{L}_{\text{Match}}$
        \STATE Update $\phi, \psi$
    \ENDFOR
\ENDWHILE
\STATE \textbf{Output}:
\STATE \quad Autoencoder $f_\theta^{\text{enc}}$ and $f_\theta^{\text{dec}}$,
\STATE \quad Embedding space dynamical function $\mathbf{v}_\phi$ and $g_\psi$.
\end{algorithmic}
\end{algorithm}

\subsection{Evaluation Metrics}
Besides the visual judgment, we employ several quantitative metrics to assess the inferred embeddings. In the simulated data, where the ground truth is known, we directly calculate the velocity accuracy (VA) between the estimated velocity and the true values. For the real datasets without ground truth, we develop two consistency metrics: velocity consistency (VC) and temporal consistency (TC) as follows.

\noindent\textbf{Velocity Accuracy (VA)}
With known ground truth, we assess predictive accuracy by computing the distance correlation (dCor) between the predicted velocity 
$\{\mathbf{v}_i^\text{pred}\}_{i=1}^N$ 
in the cell embedding space and the ground truth velocity 
$\{\mathbf{v}_i^\text{real}\}_{i=1}^N$ 
in the same space:
\begin{equation*}
   \text{VA} = \text{dCor}(\mathbf{v}^\text{real}, \mathbf{v}^\text{pred}),
\end{equation*}
where the predicted velocity $\mathbf{v}^\text{pred}$ can be directly derived from $\mathbf{v}_\phi$, and the real value $\mathbf{v}^\text{real}$ can be calculated by using coordinate transformation formula.

\noindent\textbf{Velocity Consistency (VC)}~
We propose velocity consistency (VC) to assess velocity coherence in the embedded latent space. A higher VC is expected since it indicates a smoother velocity field which boosts better robustness in downstream analysis. Specifically, the neighborhood of each cell is first identified, and then an individual average velocity similarity and overall velocity consistency are computed:
\begin{equation*}
    \text{VC}_r = \cfrac{\sum\limits_{i,j} \text{vc}_r(\mathbf{z}_i^j)}{\sum\limits_i \, N_i},\ \ 
    \text{vc}_r(\mathbf{z}_i^j) = \cfrac
    {\sum\limits_{\mathbf{z}_k^l\in \mathcal{N}_r(\mathbf{z}_i^j)} \!\!\!\! \cos \langle \mathbf{v}(\mathbf{z}_k^l), \mathbf{v}(\mathbf{z}_i^j) \rangle}
    {{|\{\mathbf{z}_k^l \in\mathcal{N}_r(\mathbf{z}_i^j) \}|}}
\end{equation*}
where  $\mathcal{N}_r(\mathbf{z}_i^j)$ indicates the r-distance neighborhood of $\mathbf{z}_i^j$, and $\mathbf{v}(\mathbf{z}_k^l)$ is the velocity vector of $\mathbf{z}_i^j$ in the embedding space. In the numerical experiment, we set $r = 0.05$.

\noindent\textbf{Temporal Consistency (TC)}~
We propose temporal consistency (TC) to assess the degree of temporal separation in the embedding space. For each cell from time point $i$, we calculate the proportion of temporally matched neighbors, namely the fraction of cells within its neighborhood that also belong to time point $i$. TC is then computed as the average of these proportions across all cells from all time points, providing a global measure of temporal coherence in the embedding:
\begin{equation*}
    \text{TC}_r =  \cfrac{\sum_{i,j} \, \text{tc}_r(\mathbf{z}_i^j)}{\sum_i \, N_i}, \ \ 
    \text{tc}_r(\mathbf{z}_i^j) = \cfrac    
 {{|\{\mathbf{z}_k^l \in\mathcal{N}_r(\mathbf{z}_i^j) \bigcap \mathbf{Z}_i\}|}}
    {{|\{\mathbf{z}_k^l \in\mathcal{N}_r(\mathbf{z}_i^j) \}|}}.
\end{equation*}

\section{Experiments}
\subsection{Experimental Setups}

\noindent\textbf{Datasets.}
We first benchmarked CellStream on the simulated data to demonstrate its capability in capturing cellular dynamics and its effectiveness of denoising in embedding construction. The cellular dynamics simulation involves a cell population with three time points and 6 genes, which gradually diverges into two cell subpopulations over time. Subsequently, three publicly available time-series datasets were utilized: a single cell RNA sequencing (scRNA-seq) dataset, a quantitative PCR (qPCR) dataset and a spatial transcriptome (ST) dataset. The EMT dataset describes an A549 cancer cell line treated with TGFB1 to induce the epithelial-mesenchymal transition (EMT) response~\cite{EMT}. The iPSC dataset captures the differentiation of induced pluripotent stem cells (iPSCs) into specific cell types, featuring a bifurcation event~\cite{iPSC}. Finally, the mouse organogenesis spatiotemporal transcriptomic atlas (MOSTA) dataset illustrates the spatiotemporal transcriptomic dynamics during organogenesis in C57BL/6 mouse embryos~\cite{MOSTA}. The data preprocessing details are left in Appendix A.2.


\noindent\textbf{Baselines and benchmarks.}
To compare the embeddings, we use two quantitative metrics, velocity consistency (VC) and temporal consistency (TC), to evaluate the performance of CellStream compared with other mainstream methods. Since there are no previous approaches for dynamics-informed embeddings based on unbalanced dynamical OT, we compare our algorithm with 6 baseline methods which can be divided into two categories: 1) unbalanced dynamical OT model, TIGON, coupled with four geometric embeddings including PCA, t-SNE, UMAP and diffusion maps~\cite{TIGON}; 2) dynamical embeddings generated by VeloViz (RNA velocity based) and MIOFlow (dynamical OT based)~\cite{VeloViz,MIOFlow}. Since RNA velocity analysis requires unspliced counts which is not available in some single-cell datasets, we use the velocity field estimated by TIGON as the input of VeloViz to approximate RNA velocity.

\begin{table*}[htbp]
\centering
\fontsize{9pt}{\baselineskip}\selectfont
\resizebox{0.9\textwidth}{!}{%
    \setlength\tabcolsep{2pt}
    \renewcommand\arraystretch{1.0}
    \begin{tabular}{c*{14}{>{\centering\arraybackslash}p{2.5em}}}
    \toprule
    \multirow{2}{*}{Method} & \multicolumn{2}{c}{\multirow{2}{*}{\textbf{CellStream(ours)}}} & \multicolumn{2}{c}{\multirow{2}{*}{VeloViz}} & \multicolumn{2}{c}{\multirow{2}{*}{MIOFlow}} & \multicolumn{8}{c}{TIGON} \\
    \cline{8-15}
    & \multicolumn{2}{c}{} & \multicolumn{2}{c}{} & \multicolumn{2}{c}{} & \multicolumn{2}{c}{PCA} & \multicolumn{2}{c}{t-SNE} & \multicolumn{2}{c}{UMAP} & \multicolumn{2}{c}{Diffusion Maps} \\
    \midrule
    dataset & VC & TC & VC & TC & VC & TC & VC & TC & VC & TC & VC & TC & VC & TC \\
    \midrule
    EMT & $\textbf{0.97}$ & $\textbf{0.99}$ & $0.88$ & $0.70$ & $0.96$ & $0.77$ & $0.66$ & $0.59$ & $0.70$ & $0.94$ & $0.67$ & $0.91$ & $0.68$ & $0.57$ \\
    iPSC & $0.97$ & $0.91$ & $0.41$ & $0.92$ & $\textbf{0.98}$ & $0.92$ & $0.32$ & $0.94$ & $0.74$ & $\textbf{0.95}$ & $0.82$ & $0.93$ & $0.78$ & $0.84$ \\
    MOSTA & $\textbf{0.98}$ & $\textbf{0.99}$ & $0.83$ & $0.89$ & $0.42$ & $0.92$ & $0.85$ & $0.99$ & $0.43$ & $0.99$ & $0.47$ & $0.99$ & $0.98$ & $0.99$ \\
    \bottomrule
    \end{tabular}
}
\caption{Performance comparison of different trajectory inference methods on three real datasets. VC indicates velocity consistency and TC indicates temporal consistency.}
\label{table:method_comparison}
\vspace{-0.3cm}
\end{table*}

\noindent\textbf{Implementation Details.}
CellStream is constructed with PyTorch 2.6.0 and Python 3.12.6. The autoencoder is set as an encoder-decoder architecture, comprising 3 hidden layers each with a dimension of 10 and RELU activations. The output dimension of the encoder is set to 2. The velocity network and growth network both employ a hidden dimension of 10 with Tanh activations, differing in layer number with 4 and 3 hidden layers respectively. The loss weights $\{{\lambda_{\text{AE}}, \lambda_{\text{WFR}}, \lambda_{\text{Match}}, \lambda_{\text{OT}}, \lambda_{\text{Mass}}}\}$ are set to $\{10,1,5,1,1\}$. $\alpha$ and $r$ are set to 1 and 0.05 respectively. Our model is optimized using equation \eqref{eq: loss function} with the Adam algorithm. The parameters of baseline methods remain as the default.

\subsection{Simulated Data with Bifurcation}
\begin{figure}[!htb]
\centering
\includegraphics[width=1\linewidth]{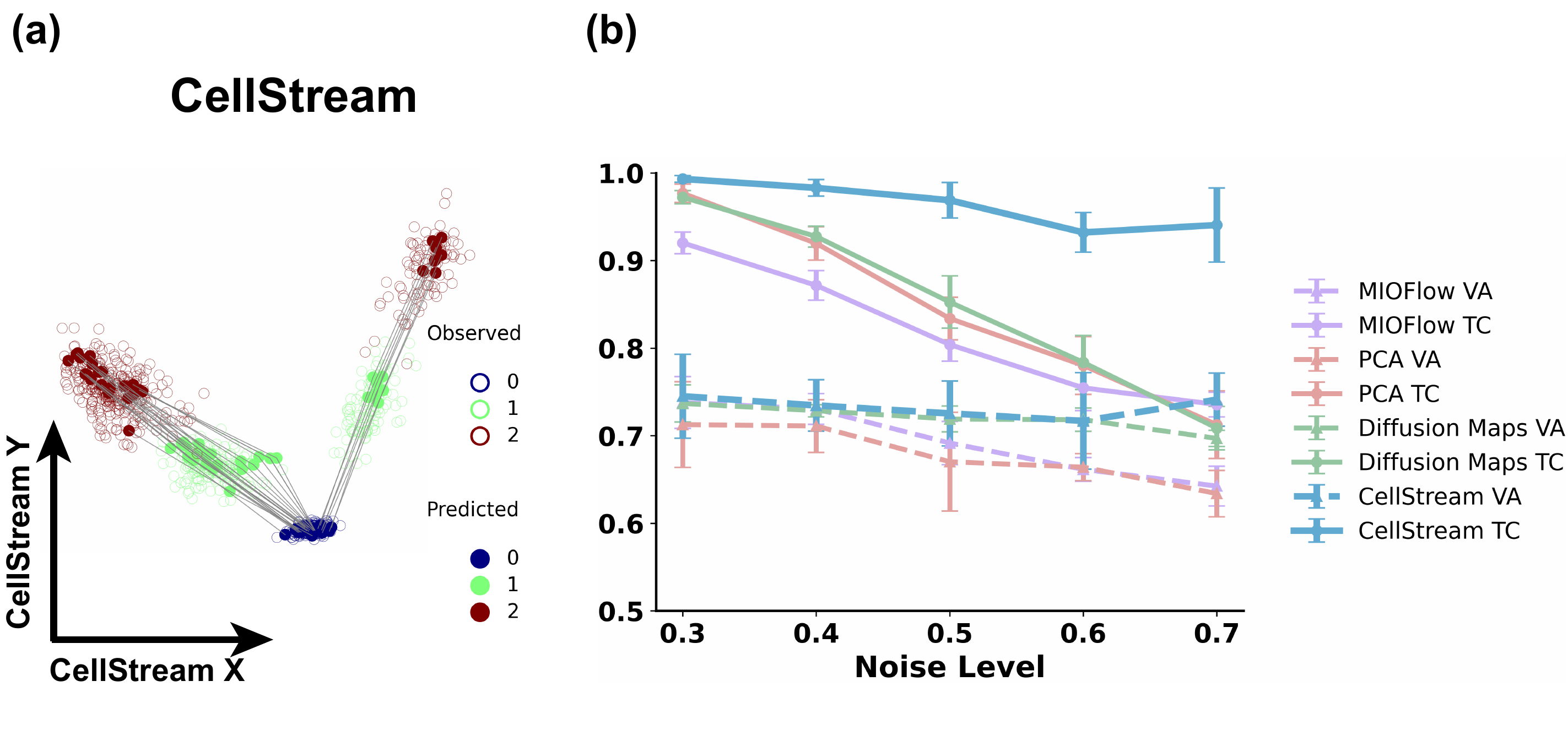}
\vspace{-0.3cm}
\caption{Application in the Simulated data. (a) The dynamics-informed embedding learned by CellStream. (b) Velocity accuracy and temporal consistency of embeddings from different methods across six runs on simulated data with increasing noise level.}
\label{sim}
\vspace{-0.5cm}
\end{figure}

We first demonstrated the performance of CellStream on simulated data with unbalanced bifurcating effects generated from a Stochastic Differential Equation (SDE) solver (Appendix A.1). We artificially added noise to mimic the technical noise of real snapshots data. CellStream successfully identifies two distinct cell populations and reconstructs their developmental trajectories from 3 artificial snapshots (Fig. \ref{sim}(a)). In order to demonstrate the noise robustness of CellStream, we constructed 5 sets of simulated data with an increasing magnitude of noise, and compared the denoising effect of CellStream with other baseline methods. Fig. \ref{sim}(b) shows that CellStream achieves high VA and TC despite the disturbance of high noise, while the performance of other methods is adversely affected. Though certain manifold learning embeddings like diffusion maps can attain relatively high VA scores under high noise conditions, there is a dramatic decrease in TC, indicating a challenge to interpret temporal structures accurately. A detailed comparison is provided in Appendix B.3.

\subsection{Epithelial-to-mesenchymal Transition}
\begin{figure}[!htb]
\centering
\includegraphics[width=1\linewidth]{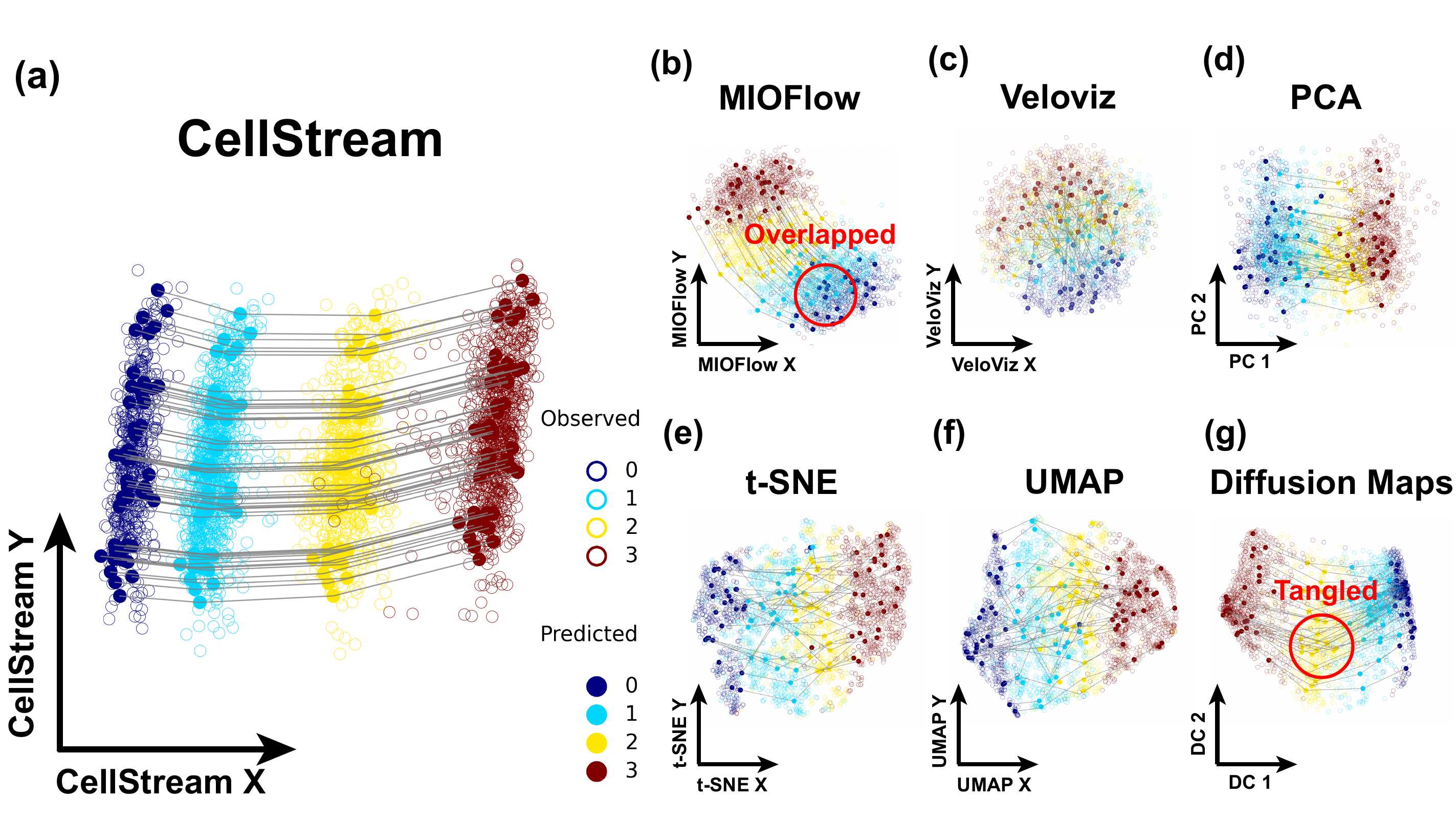}
\vspace{-0.3cm}
\caption{Application in the EMT dataset. (a) The dynamics-informed embedding learned by CellStream. (b) The embeddings learned by MIOFlow. (c) The embedding learned by VeloViz. (d-g) The embeddings from TIGON with PCA, t-SNE, UMAP and diffusion maps respectively.}
\label{emt}
\vspace{-0.5cm}
\end{figure}

We next applied our method to the epithelial-mesenchymal transition (EMT) dataset from an A549 cancer cell line treated with TGFB1~\cite{EMT} sampled across 4 time points.  In Fig. \ref{emt}(a), CellStream embeddings represent a clear cell-state transition temporal structure with highest VC and TC, aligned with the prior knowledge of EMT. In contrast, cell populations sampled from different time points overlap with each other in methods like VeloViz and MIOFlow (Fig. \ref{emt}(b)(c)). While the embeddings generated by t-SNE or UMAP informed TIGON also capture such structure, their performances are inferior to CellStream in velocity consistency (Fig. \ref{emt}(e)(f), Table \ref{table:method_comparison}). CellStream can also capture the growth dynamics consistent with relative cell population changes at each time point (Appendix B.2). 

\subsection{Induced Pluripotent Stem Cells}
\begin{figure}[!htb]
\centering
\includegraphics[width=1\linewidth]{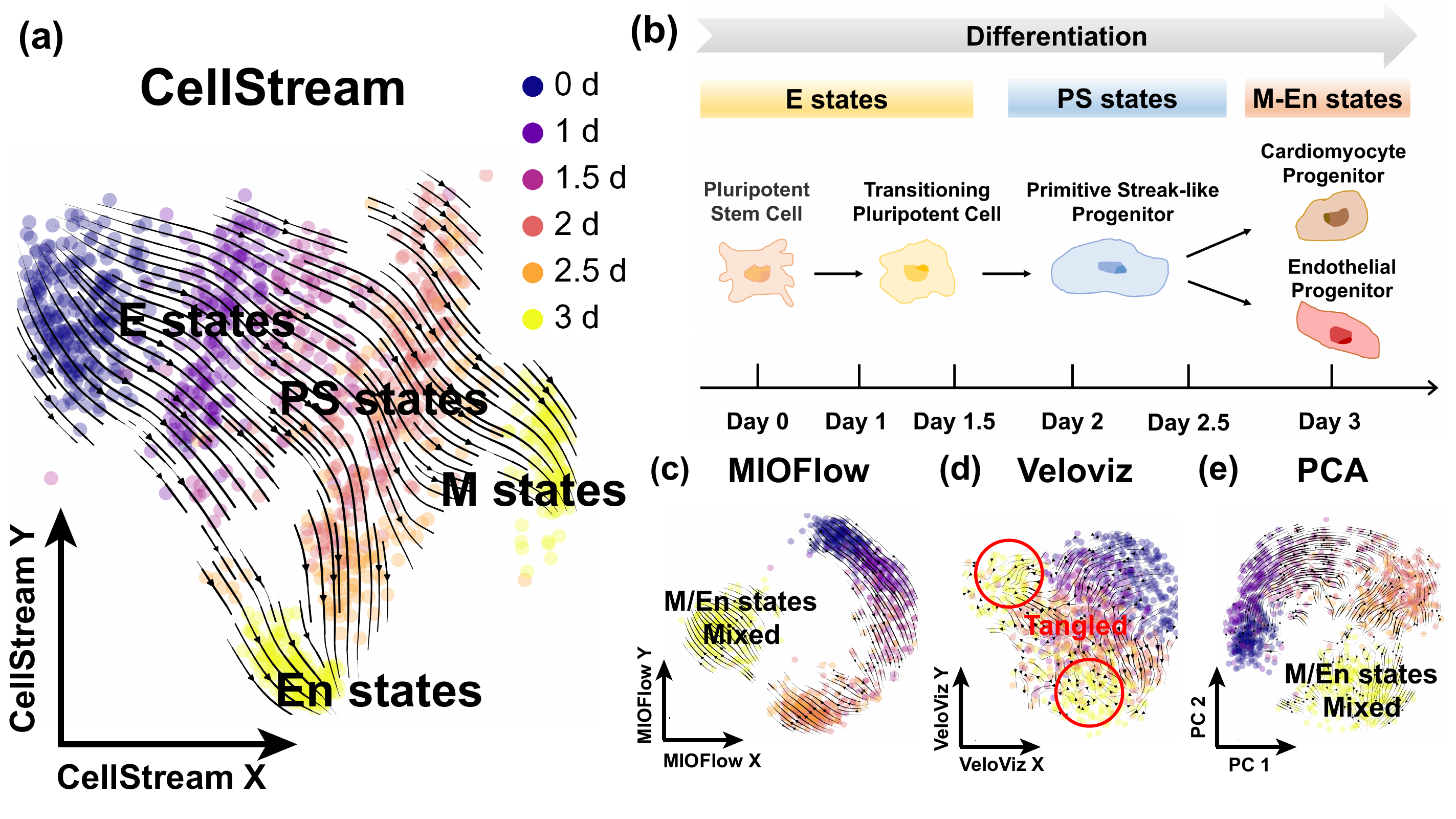}
\vspace{-0.3cm}
\caption{Application in the iPSC dataset. (a) The dynamics-informed embedding learned by CellStream. (b) Illustration of the bifuraction event in the iPSC dataset. (c) The embedding learned by MIOFlow. (d) The embedding learned by VeloViz. (e) The embedding learned by TIGON with PCA. }
\label{ipsc}
\vspace{-0.5cm}
\end{figure}

We then used the induced pluripotent stem cells (iPSCs) differentiation dataset to validate the performance of CellStream in bifurcation event~\cite{iPSC}. Day 3 is the branching point when cells initially in PS states segregated into Mesodermal (M) states or Endodermal (En) states.  We selected data at the first 6 time points that included the crucial bifurcation process. CellStream successfully reveals the cellular differentiation towards mesodermal or endodermal lineages which is not fully resolved by MIOFlow and TIGON with PCA (Fig. \ref{ipsc}(a)(c)(e)). While VeloViz embeddings represent the bifurcation event, the velocity field is tangled at the branching point (Fig. \ref{ipsc}(d)) and so do the other three embeddings(Appendix B.1).

\subsection{Spatiotemporal Analysis}

\begin{figure}[!htb]
\centering
\includegraphics[width=1\linewidth]{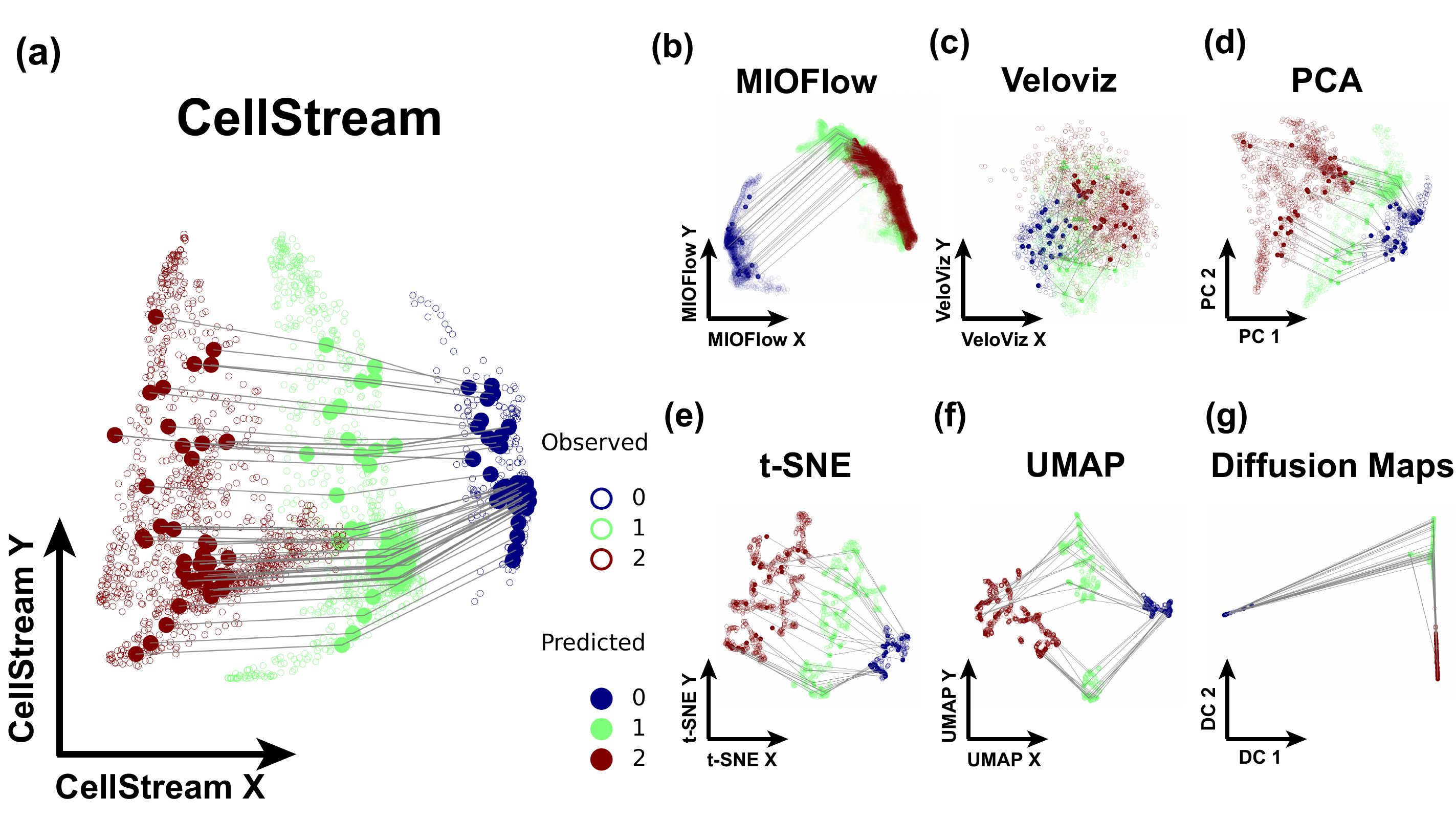}
\vspace{-0.3cm}
\caption{Application in the MOSTA dataset. (a) The dynamics-informed embedding learned by CellStream. (b) The embedding learned by MIOFlow. (c) The embedding learned by VeloViz. (d-g) The embeddings from TIGON with PCA, t-SNE, UMAP and diffusion maps respectively.}
\label{mosta}
\vspace{-0.5cm}
\end{figure}

Recently, the emergence of spatial transcriptomics (ST) technologies has enabled us to investigate the spatial distribution of gene expression. We further used the MOSTA (mouse organogenesis) dataset to assess the applicability of our method in ST data ~\cite{MOSTA}. We applied a pre-trained GNN ~\cite{STAGATE} to integrate the gene expression matrix with spatial relationships between cells, using the resulting 50-dimensional space as the input of our model. In CellStream embedding (Fig. \ref{mosta}(a)), the cell population grew dramatically at the first time point and drifted forward consistently, agreeing with the prior spatiotemporal transcriptomic atlas (Appendix B.2). However, embeddings generated by MIOFlow and VeloViz do not accurately reflect the temporal structure. The four low-dimensional manifold embeddings also exhibit incoherent and disordered results compared with CellStream (Table \ref{table:method_comparison}, Fig. \ref{mosta}(d-g)).

\subsection{Ablation studies and hyperparameter sensitivity}
We conduct ablation experiments to evaluate the contribution of different components. Specifically, we first decouple AE from the dynamics model to assess the necessity of jointly learning embeddings and cellular dynamics. Next, we dismiss the growth term to demonstrate the impact of unbalanced dynamical OT. Ablation study results are provided in Appendix B.4. Furthermore, we examine CellStream's sensitivity to parameter variance by tuning hyperparameters including $\alpha$ , $\lambda_{\text{AE}}$, $\lambda_{\text{WFR}}$, $\lambda_{\text{Match}}$, $\lambda_{\text{OT}}$,  $\lambda_{\text{Mass}}$ and $r$. We include the detailed parameter sensitivity in Appendix B.5.

\section{Conclusion}
In this paper, we introduce CellStream, a new algorithm that jointly learns low-dimensional embeddings and continuous cellular stream from static, sparse snapshots data. We have demonstrated on both simulated data and real snapshots data that the dynamical embeddings derived from CellStream yield enhanced temporal coherence and improved denoising effect in revealing the underlying cellular dynamics. However, a key limitation of CellStream is that a sufficiently potent decoder architecture is yet to be developed to project dynamics back into the original gene expression space with high reconstruction fidelity. Future directions also involve providing a deeper insight into gene regulation networks from embedding space and extending the applications to various biological scenarios, like multi-omics data or cell-to-cell communication \cite{DeepRUOTv2}. 

\section{Acknowledgments}
This work was supported by the National Natural Science Foundation of China (NSFC No. 12288101, 8206100646, T2321001 to P.Z.) and The Fundamental Research Funds for the Central Universities, Peking University. We acknowledge the support from the High-performance Computing Platform of Peking University for computation.

\bibliography{aaai2026}

\appendix
\setcounter{secnumdepth}{2} 

\section{Experiment Details}

\subsection{Simulated Data}

To evaluate the accuracy of CellStream in both dimensionality reduction and dynamical inference, we construct simulated datasets representing multi-cellular populations under multi-factorial dynamics. In this setting, the position of a cell in the gene expression space is jointly influenced by directed velocity, population-level growth, and stochastic diffusion.  

Specifically, we consider $K$ cell populations in a $D$-dimensional space ($K<D$), each with a velocity $\mathbf{v}_k(t, \mathbf{x})$, a growth rate $g_k(t, \mathbf{x})$, and a diffusion matrix $\mathbf{\Sigma}_k$ 
($k = 1, 2, \cdots, K$).  
We aggregate the velocities and growth rates as  
\begin{equation*}
\begin{split}
\mathbf{v} &= [\mathbf{v}_1; \mathbf{v}_2; \cdots; \mathbf{v}_K] \in \mathbb{R}^{D \times K},\\
\mathbf{g} &= (g_1, g_2, \ldots, g_K) \in \mathbb{R}^{K}.
\end{split}
\end{equation*}

Each population evolves according to its own velocity, growth, and diffusion dynamics. For the $k$-th population, the stochastic process satisfies  
$$
\mathrm{d} \mathbf{X}^{(k)} = \mathbf{v}_k \,\mathrm{d} t + \mathbf{\Sigma}_k \,\mathrm{d} \mathbf{W}_t,
\quad 
\text{with birth/death rate } g_k(t, \mathbf{x}),
$$
where $\mathbf{W}_t$ is a standard Wiener process.  

In our experiments, we set the gene expression space dimension to $D=6$ and the number of cell populations to $K=2$.  
The velocity matrix $\mathbf{v}$ is chosen as the first $K$ columns of the $D \times D$ identity matrix $\mathbf{I}_D$, so that the $k$-th population moves with unit speed along the positive $k$-th coordinate axis.  
The growth vector is set to $\mathbf{g} = [0.5,\, 0]$.  
The diffusion matrices are defined as  
\[
\mathbf{\Sigma}_k = \mathrm{diag}(\boldsymbol{\sigma}_k), 
\quad
\boldsymbol{\sigma}_k^i =
\begin{cases}
\sigma_{\text{main}}, & i = k,\\
\sigma_{\text{perp}}, & i \neq k,
\end{cases}
\]
where $\sigma_{\text{main}} = 0.3$ and $\sigma_{\text{perp}} = 0.1$.  
This configuration implies that each cell population exhibits a larger diffusion coefficient along its velocity direction, while sharing a uniform and smaller diffusion coefficient in the perpendicular directions, which is consistent with observations from real transcriptomic data.


For computing the velocity accuracy (VA), we need the ground-truth velocities and growth rates, denoted by $\mathbf{v}^{\text{real}}$ and $g^{\text{real}}$, in the latent space encoded by a given encoder $f^{\text{enc}}$.  
Clearly, $g^{\text{real}} = g$, whereas $\mathbf{v}^{\text{real}}$ is obtained through a coordinate transformation:
\begin{equation}
\label{eq: latent real velocity}
    \mathbf{v}^{\text{real}}(\mathbf{z}) 
    = \frac{\mathrm{d}\mathbf{z}}{\mathrm{d}t} 
    = \mathbf{J}_{f^{\text{enc}}}(\mathbf{x})\, \mathbf{v}(\mathbf{x}),
\end{equation}
where $\mathbf{J}_{f^{\text{enc}}}(\mathbf{x})$ denotes the Jacobian matrix of $f^{\text{enc}}$ with respect to $\mathbf{x}$.

\subsection{scRNA-seq Dataset Pre-processing}
We demonstrated the performance of CellStream on three publicly available temporal single-cell RNA sequencing (scRNA-seq) datasets: Epithelial-to-mesenchymal Transition (EMT) dataset,  Induced Pluripotent Stem Cells (iPSC) dataset and mouse organogenesis spatiotemporal transcriptomic atlas (MOSTA) dataset. All datasets were preprocessed before embedding construction to ensure training robustness. Owing to the overwhelming cell population of MOSTA, we applied the stratified time point sampling strategy from scIMF~\cite{scIMF} to retain 10\% of its original cell counts. 

For each dataset, we first leveraged count normalization followed by log-transformation to remove cell-specific bias and stabilize count variance. Next, we identified 3000 highly variable genes (HVGs) of the EMT and MOSTA dataset respectively. HVGs is a subset of genes that account for the majority of cell-to-cell variance, routinely used to reduce technical noise and computational costs~\cite{scIMF,scNode}. Since the iPSC dataset has already undergone gene filtering and quality control, all 96 genes involved are directly used without HVGs detection~\cite{iPSC}. Inspired by DeepRUOT, we further employ Principal Component Analysis (PCA) to reduce the dimensionality of the EMT dataset to 10, used as the input for CellStream~\cite{DeepRUOT}. For the MOSTA dataset, we additionally employed STAGATE integrate the gene expression matrix and spatial coordinate into a 50-dimensional latent space~\cite{STAGATE}. The latent representatives were subsequently used as the input features of CellStream. To ensure a clear representation and interpretation, the embedding dimension was set as 2 in all subsequent analyses. To simplify computation, we relabeled time points as consecutive integers starting from zero. The same data pre-processing pipeline was applied to CellStream and other baseline methods. All experiments were conducted Using an AMD Ryzen 7 5800H with Radeon Graphics processor (8 cores/16 threads, base frequency 3.20 GHz) and 16GB DDR4-3200MHz memory.

\subsection{Evaluation Metric}
Besides the visual judgment, we employ several quantitative metrics to assess the inferred embeddings. In the simulated data, where the ground truth is known, we directly calculate the velocity accuracy (VA) between the estimated velocity and the true values. For the real datasets without ground truth, we develop two consistency metrics, velocity consistency (VC) and temporal consistency (TC) as follows.

\noindent\textbf{Velocity Accuracy (VA)}~

With known ground truth, evaluating the algorithm essentially reduces to measuring the similarity between two sets of vectors: the predicted values and the true values. 
We adopt distance correlation (dCor) as the evaluation metric. 
Since the dynamics in this study are modeled in the latent space, the comparison is also conducted in the reduced latent space.

Specifically, CellStream produces a latent velocity field $\mathbf{v}_\phi$, from which the predicted velocities for all cells, 
$\mathbf{v}^\text{pred} = \{\mathbf{v}_i^\text{pred}\}_{i=1}^N$, 
are computed. 
The corresponding ground-truth latent velocities, 
$\mathbf{v}^\text{real} = \{\mathbf{v}_i^\text{real}\}_{i=1}^N$, 
are obtained via \eqref{eq: latent real velocity}. 
Here, $N$ denotes the total number of sampled cells, i.e., $N = \sum_{i=0}^{T-1} N_i$. 
Finally, the velocity accuracy (VA) is defined as
\begin{equation*}
   \text{VA} = \text{dCor}(\mathbf{v}^\text{real}, \mathbf{v}^\text{pred}),
\end{equation*}
which quantifies the correlation between the predicted and ground-truth velocities in the latent space.

\noindent\textbf{Velocity Consistency (VC)}~

The consistency of the velocity vector fields is widely used in related works to evaluate the trajectory estimation results.ScVelo pioneers the quantitative evaluation of velocity consistency by calculating the mean correlation of velocity $\mathbf{v}_i$ from cell $i$ with velocities from neighboring cells~\cite{scVelo}. However, the output of this approach is a vector, making it not applicable in across-methods comparison. Other metrics, like CBDir and ICVCoh, calculate consistency scores that underline cross-boundary transition and coherence within the same cell type respectively by computing cosine similarity~\cite{VeloAE,UniTVelo}. But the latter two metrics require the input of cell type annotation and ground-truth developmental directions, which are not provided in snapshot data. 

So we propose velocity consistency (VC) to assess velocity coherence in the embedded latent space. 
A smoother velocity field, reflected by a higher VC, is precisely what we aim for, as it contributes to improved robustness in downstream analyses.

We aim to compute this metric based on the cosine similarity of velocities between neighboring locations. 
Specifically, for each sampled point $\mathbf{z}_i^j$, we first calculate its cosine similarity with its neighboring locations:
\begin{equation*}
    \text{vc}_r(\mathbf{z}_i^j) = \cfrac
    {\sum\limits_{\mathbf{z}_k^l\in \mathcal{N}_r(\mathbf{z}_i^j)} \cos\left\langle \mathbf{v}(\mathbf{z}_k^l), \mathbf{v}(\mathbf{z}_i^j) \right\rangle}
    {{|\{\mathbf{z}_k^l \in\mathcal{N}_r(\mathbf{z}_i^j) \}|}}
\end{equation*}
where $\mathcal{N}_r(\mathbf{z}_i^j)$ indicates the $r$-distance neighborhood of $\mathbf{z}_i^j$, and $\mathbf{v}(\mathbf{z}_k^l)$ is the velocity vector of $\mathbf{z}_i^j$ in the embedding space. 
In the numerical experiment, we set $r = 0.05$. 
In the subsequent sensitivity analysis, we demonstrate that this metric is robust to the choice of $r$ (Table \ref{table:sensitivity of alpha and r}).
Finally, the overall velocity consistency is obtained by averaging the values across all locations
\begin{equation*}
    \text{VC}_r =  \cfrac1N\sum\limits_{i,j}\,\text{vc}_r (\mathbf{z}_i^j).
\end{equation*}
where $N$ denotes the total number of sampled cells, i.e., $N = \sum_{i=0}^{T-1} N_i$.

\noindent\textbf{Temporal Consistency (TC)}~
A clear visualization of cell state transitions is expected in the embedding space. But interpreting the developmental dynamics becomes challenging when data points from different time points exhibit spatial overlap. Therefore, we propose temporal consistency (TC) to assess the degree of temporal separation in the embedding space. 
For each cell at time point $t_i$, we first compute the proportion of its neighboring points that originate from the same time point $t_i$:
\begin{equation*}
    \text{tc}_r(\mathbf{z}_i^j) = \cfrac {{|\{\mathbf{z}_k^l \in\mathcal{N}_r(\mathbf{z}_i^j) \bigcap \mathbf{Z}_i\}|}}
    {{|\{\mathbf{z}_k^l \in\mathcal{N}_r(\mathbf{z}_i^j) \}|}}.
\end{equation*}
where $\mathcal{N}_r(\mathbf{z}_i^j)$ indicates the $r$-distance neighborhood of $\mathbf{z}_i^j$. 
In the numerical experiment, we set $r = 0.05$. 
In the subsequent sensitivity analysis, we demonstrate that this metric is robust to the choice of $r$ (Table \ref{table:sensitivity of alpha and r}).
The overall proportion is then obtained by averaging these values over all locations
\begin{equation*}
    \text{TC}_r =  \cfrac1N\sum_{i,j} \, \text{tc}_r(\mathbf{z}_i^j).
\end{equation*}
where $N$ denotes the total number of sampled cells, i.e., $N = \sum_{i=0}^{T-1} N_i$.

\section{Additional Results}

\subsection{iPSC embeddings}
We utilized the induced pluripotent stem cells(iPSCs) differentiation dataset to validate the performance of CellStream in a real-world bifurcation event. The iPSC dataset featured two apparent developmental trajectories at day 3. Cells initially in a progenitor state bifurcated into either a mesodermal (M) state or an endodermal (En) state at the tipping point~\cite{iPSC}. 

\begin{figure*}[!htb]
\centering
\includegraphics[width=1\linewidth]{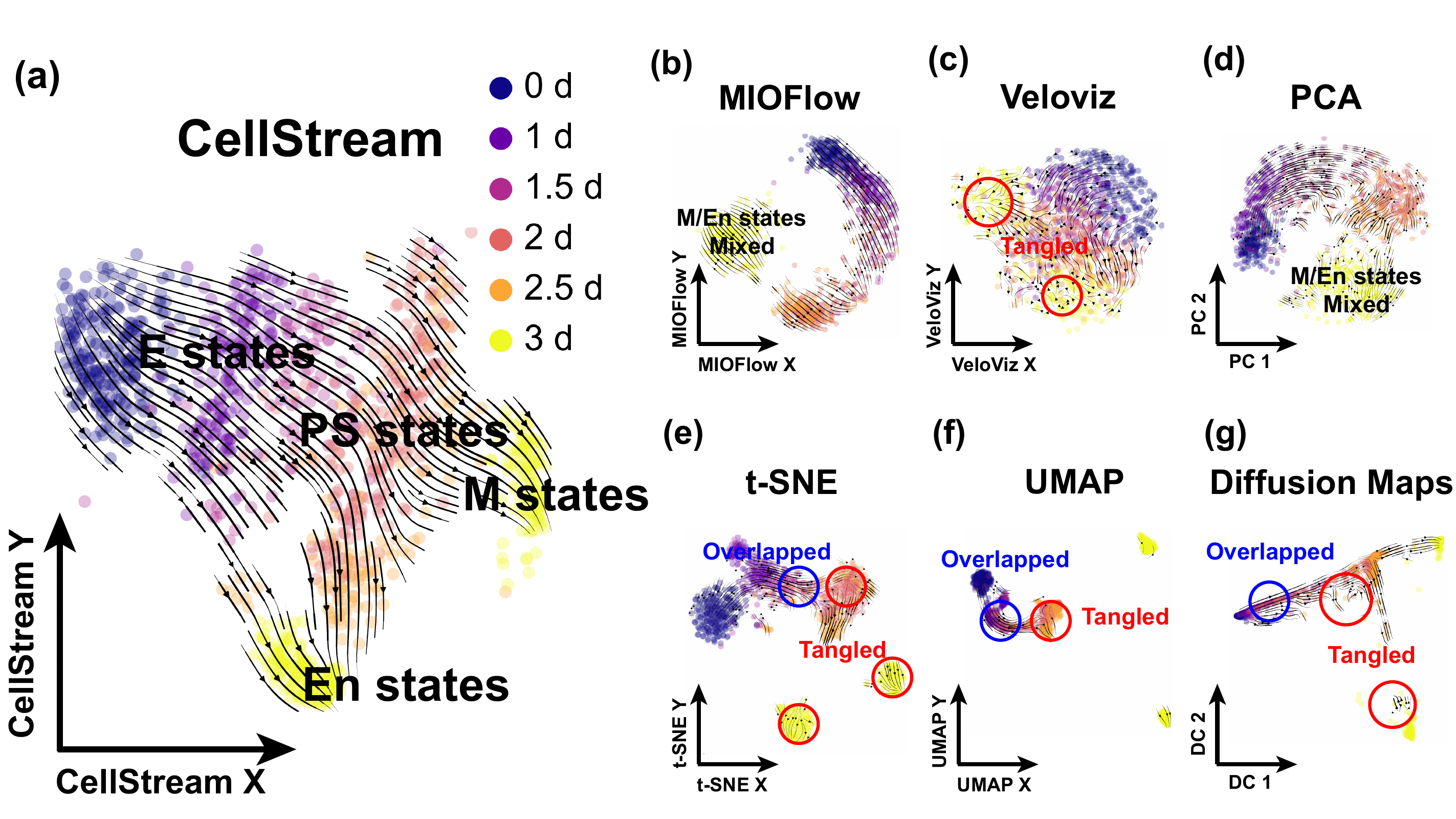}
\vspace{-0.3cm}
\caption{Application in the iPSC dataset. (a) The dynamics-informed embedding learned by CellStream. (b) The embedding learned by MIOFlow. (c) The embedding learned by VeloViz. (d-g) The embedding learned by TIGON coupled with PCA, t-SNE, UMAP and diffusion maps respectively.}
\label{ipsc_full}
\vspace{-0.5cm}
\end{figure*}

All the embeddings generated by CellStream and other 6 baseline methods are shown in Fig.\ref{ipsc_full}. CellStream successfully reveals the cellular differentiation towards mesodermal or endodermal lineages which is neglected by MIOFlow and PCA informed by TIGON (Fig.\ref{ipsc_full}(a)(b)(d)). Although the other four methods also represent the bifurcation event, cells from different time points overlap with each other, barely discernible and interpretable (Fig.\ref{ipsc_full}(c)(e)(f)(g)). What' more, their velocity field is tangled and disordered, posing a threat to downstream analysis.

\subsection{Growth learned by CellStream}
\begin{figure}[!htb]
\centering
\includegraphics[width=1\linewidth]{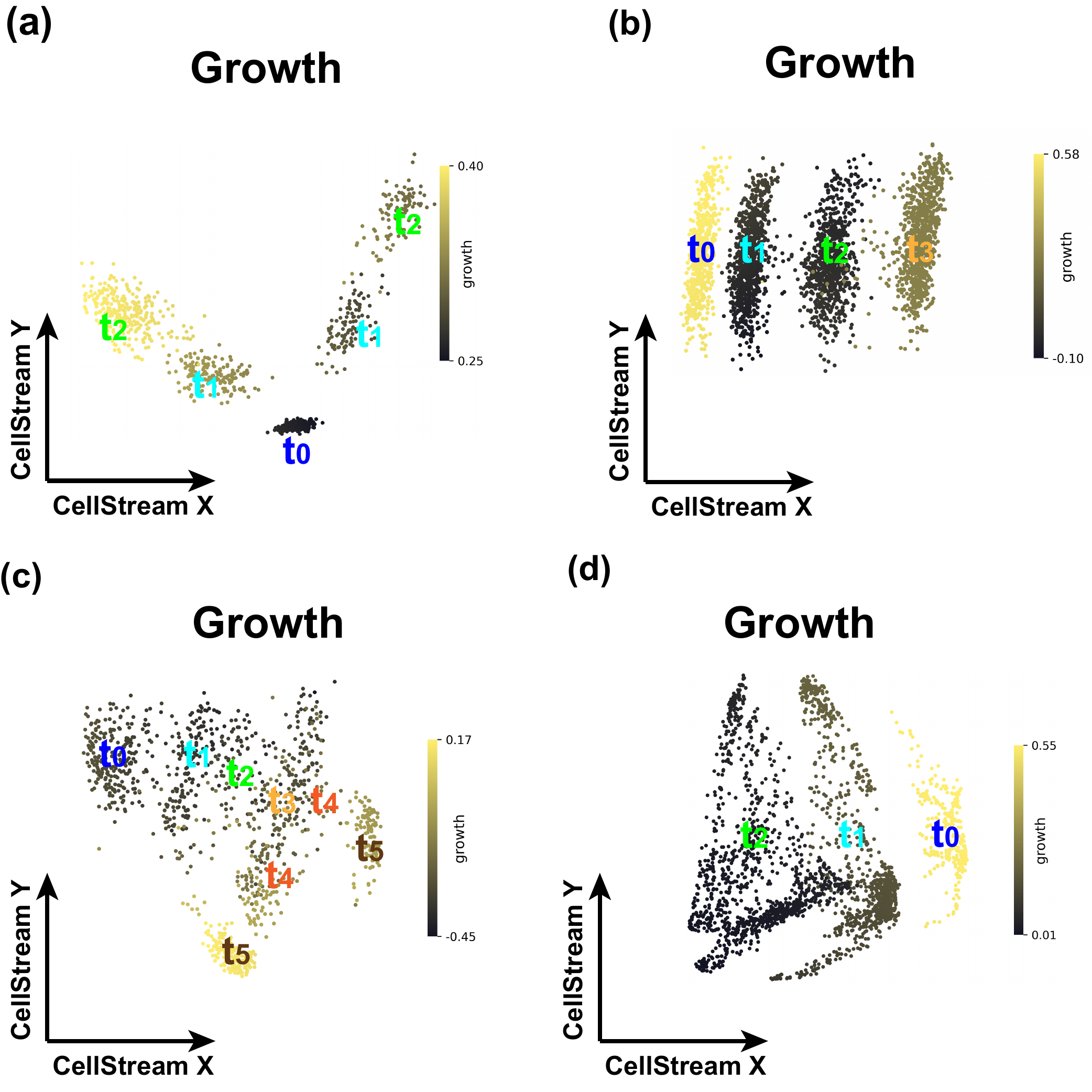}
\vspace{-0.3cm}
\caption{The cellular growth of the simulated data and three real data esitmated by CellStream. (a) Simulated data. (b) The EMT dataset. (c) The iPSC dataset. (d) The MOSTA dataset.}
\label{growth}
\vspace{-0.5cm}
\end{figure}

We also infer the cell population growth of the simulated data and the real time-series data. For simulated data and the MOSTA dataset, CellStream reveals a consistent cell proliferation over time, which corresponds with the prior knowledge (Fig.\ref{growth}(a)(d)). Though the EMT dataset and the iPSC dataset both feature more complicated growth processes as growth and death happen alternately, CellStream still infers the true growth pattern (Fig.\ref{growth}(b)(c)). Overall, CellStream is capable of recovering the growth and transitions accurately from static, sparse single-cell transcriptomics data .

\subsection{Comparison of embeddings learned from different methods across six runs on simulated data}
In order to demonstrate the noise robustness of CellStream, we constructed 5 sets of simulated data with an increasing magnitude of noise, and compared the denoising effect of CellStream with other baseline methods. Specifically, we gradually increase $\sigma_{\text{main}}$ from 0.3 to 0.7 and compare the corresponding velocity accuracy (VA) and temporal consistency (TC) of different methods. Since we construct the simulated data in the original gene expression space, we need to map the spatial distribution and growth truth velocity to the embedding space for VA calculation. However,  VeloViz, t-SNE and UMAP apply nonlinear transformations $f(\mathbf{x})$ to positions, but velocity requires $\frac{\mathrm{d}}{\mathrm{d}t}[f(\mathbf{x})] = \mathbf{J}(\mathbf{x}) \cdot \frac{\mathrm{d}\mathbf{x}}{\mathrm{d}t}$, where the position-dependent Jacobian $\mathbf{J}(\mathbf{x})$ is undefined for these embedding construction methods. Therefore, here we compare CellStream to MIOFlow, TIGON informed PCA and diffusion maps in noise sensitivity.
 
 Table~\ref{table:detailed comparison on simulation} shows that CellStream achieves high VA and TC despite the interference of high noise, while the performance of other methods is adversely affected. Though VA scores of certain manifold learning embeddings like diffusion maps can remain relatively steady under high noise conditions, there is a dramatic decrease in TC scores, neglecting temporal structure across different time points.

\begin{table*}[htbp]
\centering
\fontsize{9pt}{\baselineskip}\selectfont
\resizebox{1\textwidth}{!}{%
    \setlength\tabcolsep{2pt}
    \renewcommand\arraystretch{1.0}
    \begin{tabular}{c*{8}{>{\centering\arraybackslash}p{6em}}}
    \toprule
    \multirow{2}{*}{Method} & \multicolumn{2}{c}{\multirow{2}{*}{\textbf{CellStream(ours)}}} & \multicolumn{2}{c}{\multirow{2}{*}{MIOFlow}} & \multicolumn{4}{c}{TIGON} \\
    \cline{6-9}
    & \multicolumn{2}{c}{} & \multicolumn{2}{c}{} & \multicolumn{2}{c}{PCA} & \multicolumn{2}{c}{Diffusion Maps} \\
    \midrule
    noise level & VA & TC & VA & TC & VA & TC & VA & TC \\
    \midrule
    0.3 & $\boldsymbol{0.740{\scriptstyle\pm0.048}}$ & $\boldsymbol{0.993{\scriptstyle\pm0.004}}$ & $0.738{\scriptstyle\pm0.030}$ & $0.920{\scriptstyle\pm0.012}$ & $0.713{\scriptstyle\pm0.049}$ & $0.977{\scriptstyle\pm0.010}$ & $0.737{\scriptstyle\pm0.021}$ & $0.972{\scriptstyle\pm0.008}$ \\
    0.4 & $\boldsymbol{0.735{\scriptstyle\pm0.029}}$ & $\boldsymbol{0.983{\scriptstyle\pm0.009}}$ & $0.731{\scriptstyle\pm0.018}$ & $0.872{\scriptstyle\pm0.017}$ & $0.711{\scriptstyle\pm0.030}$ & $0.920{\scriptstyle\pm0.019}$ & $0.729{\scriptstyle\pm0.007}$ & $0.927{\scriptstyle\pm0.012}$ \\
    0.5 & $\boldsymbol{0.726{\scriptstyle\pm0.037}}$ & $\boldsymbol{0.969{\scriptstyle\pm0.023}}$ & $0.692{\scriptstyle\pm0.025}$ & $0.800{\scriptstyle\pm0.019}$ & $0.670{\scriptstyle\pm0.056}$ & $0.834{\scriptstyle\pm0.024}$ & $0.719{\scriptstyle\pm0.015}$ & $0.853{\scriptstyle\pm0.030}$ \\
    0.6 & $0.717{\scriptstyle\pm0.055}$ & $\boldsymbol{0.932{\scriptstyle\pm0.023}}$ &  $0.662{\scriptstyle\pm0.014}$ & $0.755{\scriptstyle\pm0.026}$ &  $0.664{\scriptstyle\pm0.015}$ & $0.780{\scriptstyle\pm0.033}$ & $\boldsymbol{0.718{\scriptstyle\pm0.013}}$ & $0.784{\scriptstyle\pm0.031}$\\
    0.7 & $\boldsymbol{0.741{\scriptstyle\pm0.030}}$ & $\boldsymbol{0.941{\scriptstyle\pm0.042}}$ & $0.643{\scriptstyle\pm0.023}$ & $0.736{\scriptstyle\pm0.014}$ & $0.634{\scriptstyle\pm0.026}$ & $0.713{\scriptstyle\pm0.039}$ & $0.697{\scriptstyle\pm0.009}$ & $0.709{\scriptstyle\pm0.025}$\\
    \bottomrule
    \end{tabular}
}
    \caption{Detailed comparison of different trajectory inference methods on simulated data with an increasing magnitude of noise .We show the mean value with one standard deviation. VA indicates velocity accuracy and TC indicates temporal consistency.}
    \label{table:detailed comparison on simulation}
\end{table*}

\subsection{Ablation Studies}
We conducted ablation experiments to assess the significance of each component. Specifically, we examined the influence of the growth term by setting $\alpha=0$, and decoupled the autoencoder from cellular dynamics model to demonstrate the importance of interactive learning. Other training strategies remain the same except for the target components. For example, for the EMT dataset, we trained the autoencoder for 300 epochs with a learning rate of 0.001, and the cellular dynamics model for 3000 epochs with a learning rate of 0.0003. 

Table~\ref{table:ablation studies} shows that velocity accuracy (VA) and temporal consistency (TC) are subject to the absence of the growth term or interactive learning method. Overall, the joint learning of embedding space and latent dynamics based on unbalanced dynamical OT guarantees the representation of continuous streams from static snapshots of single-cell gene expression.

\begin{table*}[htbp]
\centering
\fontsize{9pt}{\baselineskip}\selectfont
\resizebox{1\textwidth}{!}{%
    \setlength\tabcolsep{2pt}
    \renewcommand\arraystretch{1.0}
    \begin{tabular}{c*{6}{>{\centering\arraybackslash}p{8em}}}
    \toprule
    Method & \multicolumn{2}{c}{\textbf{CellStream}} & \multicolumn{2}{c}{CellStream w/o decoupled AE} & \multicolumn{2}{c}{CellStream w/o growth} \\
    \cline{1-7}
    dataset & VA & TC & VA & TC & VA & TC \\
    \midrule
    EMT & $\boldsymbol{0.97}$ & $\boldsymbol{0.99}$ & $0.57$ & $0.98$ & $0.91$ & $0.73$ \\
    iPSC & $\boldsymbol{0.97}$ & $\boldsymbol{0.91}$ & $0.70$ & $0.88$ & $0.80$ & $0.87$ \\
    MOSTA & $0.98$ & $\boldsymbol{0.99}$ & $0.98$ & $0.84$ & $\boldsymbol{0.99}$ & $0.97$ \\
    \bottomrule
    \end{tabular}
}
    \caption{Ablation studies on 3 time-series snapshot data.}
    \label{table:ablation studies}
\end{table*}

\subsection{Hyperparameter Sensitivity}
We evaluated CellStream's performance using different hyperparameter settings to assess its robustness to parameters. First, we tested the effect of loss weights. We used 6 different loss weight settings and calculated VC and TC with other parameters mixed. As shown in Table~\ref{table:loss weights sensitivity}, CellStream achieves high VC and TC in all three real scRNA-seq datasets in different loss weight combinations. This implies CellStream's robustness to loss weights.

\begin{table*}[htbp]
\centering
\fontsize{9pt}{\baselineskip}\selectfont
\resizebox{1\textwidth}{!}{%
    \setlength\tabcolsep{2pt}
    \renewcommand\arraystretch{1.0}
    \begin{tabular}{c*{12}{>{\centering\arraybackslash}p{4em}}}
    \toprule
    \multirow{2}{*}{Loss weights} & \multicolumn{12}{c}{$\{{\lambda_{\text{AE}}, \lambda_{\text{WFR}}, \lambda_{\text{Match}}, \lambda_{\text{OT}}, \lambda_{\text{Mass}}}\}$} \\
    \cline{2-7}\cline{8-13}
    & \multicolumn{2}{c}{$\{\boldsymbol{{10,1,5,1,1}\}}$} & \multicolumn{2}{c}{$\{{8,1,5,1,1}\}$} & \multicolumn{2}{c}{$\{{10,1,4,1,1}\}$} & \multicolumn{2}{c}{$\{{10,0.8,5,1,1}\}$} & \multicolumn{2}{c}{$\{{10,1,5,1.5,1}\}$} & \multicolumn{2}{c}{$\{{10,1,5,1,1.5}\}$} \\
    \midrule
    dataset & VC & TC & VC & TC & VC & TC & VC & TC & VC & TC & VC & TC \\
    \midrule
    EMT & $0.97$ & $0.99$ & $0.92$ & $0.99$ & $0.93$ & $0.97$ & $0.97$ & $0.99$ & $0.93$ & $0.96$ & $0.95$ & $0.98$ \\
    iPSC & $0.97$ & $0.91$ & $0.94$ & $0.91$ & $0.92$ & $0.88$ & $0.97$ & $0.90$ & $0.97$ & $0.90$ & $0.97$ & $0.88$ \\
    MOSTA & $0.98$ & $0.99$ & $0.99$ & $0.97$ & $0.98$ & $0.97$ & $0.99$ & $0.97$ & $0.97$ & $0.98$ & $0.97$ & $0.99$ \\
    \bottomrule
    \end{tabular}
}
    \caption{Results of different loss weights settings on 3 temporal scRNA-seq datasets. We change the 5 loss weights respectively with other hyperparameters fixed to demonstrate the loss weights sensitivity.}
    \label{table:loss weights sensitivity}
\end{table*}

Then we applied different $\alpha$ values, $\{0.8,1,1.2\}$ and calculated the corresponding VC and TC. We can observe from Table.\ref{table:sensitivity of alpha and r} that quantitative metrics are sensitive to $\alpha$ value. The conclusion is consistent with ablation studies as $\alpha$ value controls the relative weight of the velocity term and growth term which is critical in trajectory inference. We also varied $r$ from $\{0.01,0.05,0.1\}$ when computing VC and TC. CellStream boosts high VC and TC despite different $r$ values, which suggests the inherently enhanced velocity coherence and temporal coherence of CellStream embeddings (Table.\ref{table:sensitivity of alpha and r}).

\begin{table*}[htbp]
\centering
\fontsize{9pt}{\baselineskip}\selectfont
\resizebox{1\textwidth}{!}{%
    \setlength\tabcolsep{2pt}
    \renewcommand\arraystretch{1.0}
    \begin{tabular}{c*{12}{>{\centering\arraybackslash}p{4em}}}
    \toprule
    \multirow{2}{*}{Hyperparameter} & \multicolumn{6}{c}{$\alpha$} & \multicolumn{6}{c}{$r$} \\
    \cline{2-7}\cline{8-13}
    & \multicolumn{2}{c}{$\boldsymbol{\alpha = 1}$} & \multicolumn{2}{c}{$\alpha = 0.8$} & \multicolumn{2}{c}{$\alpha = 1.2$} & \multicolumn{2}{c}{$\boldsymbol{r = 0.05}$} & \multicolumn{2}{c}{$r = 0.01$} & \multicolumn{2}{c}{$r = 0.1$} \\
    \midrule
    dataset & VC & TC & VC & TC & VC & TC & VC & TC & VC & TC & VC & TC \\
    \midrule
    EMT & $0.97$ & $0.99$ & $0.84$ & $0.99$ & $0.86$ & $0.99$ & $0.97$ & $0.99$ & $0.98$ & $0.99$ & $0.96$ & $0.92$ \\
    iPSC & $0.97$ & $0.91$ & $0.92$ & $0.86$ & $0.93$ & $0.83$ & $0.97$ & $0.91$ & $0.96$ & $0.97$ & $0.96$ & $0.84$ \\
    MOSTA & $0.98$ & $0.99$ & $0.98$ & $0.97$ & $0.98$ & $0.97$ & $0.98$ & $0.99$ & $0.99$ & $0.99$ & $0.99$ & $0.98$ \\
    \bottomrule
    \end{tabular}
}
    \caption{Hyperparameter sensitivity of $\alpha$ and $r$ on 3 time-series snapshot data. The default settings of $\alpha$ and $r$ in CellStream is 1 and 0.05 respectively. Here, we change $\alpha$ to 0.8 and 1.2, $r$ to 0.01 and 0.1 and compare the correspond velocity consistency (VC) and temporal consistency (TC).}
    \label{table:sensitivity of alpha and r}
\end{table*}

\end{document}